\documentclass[aps,pra,
preprint
]{revtex4}
\def\frac#1#2{{\textstyle{#1\over#2}}} 

\def\ket#1{| #1\rangle}

\def\ul#1{\underline{#1}}

\def\remark{{\bf Remark} \qquad}
\def\proof{{\bf Proof} \qquad}

\def\R{\hbox{\rm I \kern-5pt R}}

\usepackage{amssymb}
\usepackage{amsmath}
\usepackage[sort&compress]{natbib}
\usepackage[dvips]{graphicx}

\begin{document}


\title{Secure Classical Bit Commitment using Fixed 
Capacity Communication Channels}
\author{ Adrian Kent}
\email{a.p.a.kent@damtp.cam.ac.uk}
\affiliation{Centre for Quantum Computation,
Department of Applied Mathematics and
Theoretical Physics, University of Cambridge,
Wilberforce Road, Cambridge CB3 0WA, U.K.}
\date{ February 2005 (final version) }

\begin{abstract}
If mutually mistrustful parties A and B control two or more appropriately
located sites, special relativity can be used to guarantee that a pair
of messages exchanged by A and B are independent.  
In earlier work, we used this fact to define a
relativistic bit commitment protocol, RBC1, in which security
is maintained by exchanging a sequence of messages
whose transmission rate increases exponentially in time.  We define
here a new relativistic protocol, RBC2, which requires only a constant
transmission rate and could be practically implemented.  We prove that
RBC2 allows a bit commitment to be indefinitely maintained with
unconditional security against all classical attacks.  
We examine its security against quantum attacks, and show that it is 
immune from the class of attacks shown by Mayers and Lo-Chau to render 
non-relativistic quantum bit commitment protocols insecure.    
\end{abstract}
\maketitle
\section{Introduction} 

The field of quantum cryptography was opened up by   
Wiesner's early investigation \cite{wiesner} of the application of quantum
information to cryptography, and the ensuing discoveries by Bennett and
Brassard of secure quantum key distribution \cite{BBeightyfour}
and by Ekert of entanglement-based quantum key distribution \cite{ekert}.
Quantum cryptography is now a flourishing area of theoretical research.
Its successes raise a broader theoretical question: it would be very 
interesting to know precisely which cryptographic tasks,
other than key distribution, can be implemented in such a way
that their security is guaranteed by physical principles, without
any additional computational assumptions.
An important first step would be to establish precisely which
physical principles can guarantee cryptographic security in
some serious application.  

This paper focusses on one task, bit commitment, and one physical
principle, the impossibility of superluminal signalling: that is,
of sending signals faster than light speed.  We begin by briefly
explaining both.  

Roughly speaking --- precise definitions are given in
the next section --- bit commitment is the cryptographic version
of a securely sealed envelope.  In the commitment phase
of a bit commitment protocol, Alice supplies Bob with 
data that commit her to the value
of a bit, without allowing Bob to infer that value.   
In the unveiling phase, which takes place after
commitment, if and when Alice wishes, she supplies Bob with further data that
reveal the value of the bit to which she was committed.
We are particularly interested in protocols which are 
{\it unconditionally secure}, in the sense that the 
laws of physics imply that neither party can cheat, 
regardless of the technology or computing power available to them. 

The impossibility of faster than light signalling is
guaranteed if the standard understanding of causality within
Einstein's special theory of relativity is correct.  
To characterise its cryptographic relevance requires a little 
more discussion: readers without a 
background in physics may wish to skip the technical details in
the remainder of this paragraph,
which are not needed to understand the main part of the paper.  
We assume that physics takes place in flat Minkowski
spacetime, with the Minkowski causal structure.  
This is not exactly correct, of course: according to
general relativity and to experiment, spacetime is curved.  
But it is true to a good enough approximation
for any protocol implemented on or near Earth.  
In principle, the timing constraints of our protocol should
take into account the error in the approximation.  Other than
this, the known corrections arising from general relativity do not
affect our security discussion.  Discussion of the hypothetical
cryptographic relevance of other more exotic and speculative
general relativistic phenomena can be found in Ref.~\cite{kentrel}.

Nothing in science is beyond doubt, but it seems fair to say that
the impossibility of superluminal signalling provides as solid a 
foundation for a security argument as a cryptologist could wish for.   
It would take a major scientific revolution 
for our protocols to be revealed as insecure because 
special or general relativity turned out be 
incorrect --- just as it would for quantum key distribution or
other quantum cryptography protocols to be revealed as insecure
because quantum theory proved to be incorrect.  
(That said, it should be added that at the time of writing 
the parallel between the two cases is not perfect: some quantum 
key distribution protocols have been proven secure against 
eavesdroppers equipped with arbitrarily powerful quantum computers, 
while we have not so far been able to prove that our 
bit commitment protocols are secure against general quantum attack.) 

The recent history of work on secure physical implementations of bit
commitment is complex and interesting.  Initially, attention
was focussed entirely on non-relativistic quantum protocols.     
Bennett and Brassard \cite{BBeightyfour} described a simple 
quantum bit commitment protocol which is secure against 
both parties given current technology; they also pointed out 
that it would be insecure against a committer who is able to make and
store entangled singlet states.    
Some time later, Brassard, Cr\'epeau, Jozsa and Langlois (BCJL) \cite{BCJL}  
produced a quantum bit commitment protocol which, at the time, 
they claimed to have proved to be unconditionally secure against 
attacks by either party.  

The subject was then transformed by the remarkable 
and celebrated insights of Mayers and Lo-Chau.  
A detailed history would need to consider unpublished 
work and also earlier papers that were circulated or 
temporarily archived (e.g. \cite{mayersvone}). 
Those who actively participated in these developments
are best placed to supply such a history, and  
Mayers' and Lo-Chau's accounts can be found in the
references cited below.  Not having been an 
active participant, I here discuss only  
Mayers' and Lo-Chau's published or permanently 
archived papers. 

Lo and Chau \cite{lochauprl,lochau} pointed out the existence
of an inherently quantum cheating strategy which implies that 
quantum bit commitment protocols that are perfectly secure
against Bob are completely insecure against Alice.  
Essentially the same point was made by 
Mayers \cite{mayersvthree,mayersprl,mayersone} 
at around the same time.  Mayers showed further that the essential
intuition underlying this strategy can be extended to imply
that, for a large class of quantum schemes, not only 
is perfect security against both Alice and Bob impossible, 
but security in the usual cryptographic sense --- in which
non-zero cheating probabilities are tolerated provided that
they can be made arbitrarily small by adjusting a security
paramater --- is also impossible.  In particular, as Lo and
Chau argued \cite{lochauprl} and Mayers rigorously 
proved \cite{mayersvthree}, the BCJL protocol is insecure.  

Mayers \cite{mayersprl} briefly discussed the possibility of 
using relativistic signalling constraints for bit commitment,
and suggested that his version of the no-go theorem should also 
apply to relativistic protocols.
Brassard, Cr\'epeau, Mayers and Salvail (BCMS) \cite{bcms}
subsequently examined one possible strategy for bit commitment based
on temporary relativistic signalling constraints, based on an earlier protocol
of Ben-Or, Goldwasser, Kilian and Widgerson (BGKW) \cite{bgkw}, and showed that it was indeed
insecure against quantum attacks.   

In fact, though, relativistic protocols
{\it can} evade the Mayers and Lo-Chau no-go theorems \cite{kentrel}.  
Ref.~\cite{kentrel} describes a relativistic bit commitment protocol
which iteratively 
combines BGKW's bit commitment technique with relativistic signalling
constraints.   The protocol does not belong to the classes considered by Mayers 
and Lo-Chau, and their no-go results do not apply: in particular,
the model is demonstrably not vulnerable to the Mayers-Lo-Chau cheating
strategy.  It allows 
bit commitments to be maintained indefinitely, and is conjectured
to be unconditionally secure against both classical and quantum attacks.  
However, it has a serious practical weakness, as it requires the 
communication rates between the parties to
increase exponentially in time in order to sustain the
commitment \cite{kentrel}.

The main point of this paper is to describe 
a new relativistic protocol which allows 
a committer and recipient, who each control more than one
suitably located separated site, to sustain
a bit commitment indefinitely.   
The protocol requires only a constant 
rate of communication between the parties.    
The communications need to be continued indefinitely to maintain
the commitment.  It is shown that the protocol is
unconditionally secure against all classical attacks by
either party.  I conjecture that it is also unconditionally
secure against all quantum attacks.    

This represents theoretical progress, and also opens up
new possibilities for practical cryptography.  
The relativistic protocol described in Ref.~\cite{kentrel}
is impractical in a rather strong sense: the laws of physics
appear to preclude using it to maintain a commitment indefinitely, 
since current physical theories imply an upper bound on the
attainable rate of communications emanating from any finite
spatial region.  That protocol aside, all previous 
methods of bit commitment 
have been able to guarantee security only modulo the assumed
computational difficulty of some task (such as factoring a large
number) or modulo the assumed physical difficulty of some process
(such as breaking into a locked safe, or storing entangled quantum
states \cite{BBeightyfour}).  That is, all previous bit
commitment protocols have in principle been insecure against a
sufficiently technologically advanced cheater.

The new protocol described here not only has the
theoretical virtue of needing only a constant communication
rate, but also, as we explain below, 
can be implemented with current technology.  If, as we 
conjecture, the protocol is secure
against all quantum attacks as well as classical attacks, it 
gives a practical solution to the bit commitment problem
which is unconditionally secure in the sense that it relies on no computational
complexity assumptions, classical or quantum.  It relies on no
physical assumptions either, other than the well-established principle
that signals cannot travel faster than light.

For completeness, we note that Popescu \cite{popescu} subsequently
pointed out that Mayers' and Lo-Chau's models of cryptographic protocols 
neglect the existence of quantum superselection rules.
One might perhaps think this gives another way of using a
physical principle to ensure the security of a quantum bit 
commitment scheme, but this possibility was later excluded by Mayers, 
Kitaev and Preskill \cite{mkp}.  

\section{Definitions and conventions} 

For the main part of this paper, we assume that Alice's and Bob's 
actions are limited to those allowed by classical 
physics, under which heading 
we include special relativity but not general relativity nor, 
of course, quantum theory.  
This means that in particular we allow the parties in the
protocol to send, share and manipulate classical information
but not quantum information.  We are thus content for the moment
to use a definition of bit commitment that is adequate in the 
context of classical
information theory, and postpone till section 7 discussion of 
the subtleties that arise when considering quantum information.   

First, we define the general form of a classical bit commitment
protocol.  A protocol must require Alice and Bob to exchange
classical information according to prescribed rules, with probability
distributions prescribed whenever they are required to make random 
choices, and
perhaps with some prescribed constraints on the places from and times
at which the information is sent.  In Alice's case, the rules should
depend on the value of the bit $b$ to which she wishes to commit
herself.  The protocol must include a {\it commitment phase} with
a definite end, after which Alice is, if she has
followed the protocol correctly, {\it committed} to a bit $b$ of her
choice.  It may also include a {\it sustaining phase} after the
commitment, during which both parties continue communications and/or
other operations in order to 
maintain the commitment.  And it must include an {\it unveiling phase}, 
which Alice can initiate if she chooses, in which she {\it unveils} the 
committed bit to Bob by supplying
further information which is supposed to convince him that she 
was committed to the bit $b$. 

An {\it attempted unveiling of the bit $b$}
is any attempt by Alice to use the unveiling procedure, at any
point after the commitment, by sending 
information to Bob in the hope that
it constitutes satisfactory evidence (as defined by the protocol) 
that Alice was committed to 
the bit $b$.  An attempted unveiling
of the bit $b$ might fail either by producing satisfactory evidence
that Alice was actually committed to the
bit $\bar{b}$, or else by failing to produce satisfactory evidence
that she was committed to either bit value.   
A {\it successful unveiling of the bit $b$} is an attempt by
Alice to use the unveiling procedure which succeeds in producing
satisfactory evidence (as defined by the protocol) 
that she was committed to the bit $b$.  Note that
in principle this could happen even though Alice has not 
actually followed the protocol for committing and unveiling
the bit $b$. 

For the protocol to be {\it perfectly secure}
against Bob it must guarantee that,
however he proceeds, he cannot obtain any information about the committed 
bit unless and until Alice chooses to unveil it. 
For it to be {\it perfectly secure}
against Alice, it must guarantee that,
however she proceeds, after the point at which she is supposed to
be committed, either her probability of successfully unveiling $0$
is zero or her probability of successfully unveiling $1$ is zero. 
The protocol is {\it secure} against Bob if it includes some
security parameter $n$ and has the property that the expectation
value of the Shannon information Bob can obtain about 
the committed bit is bounded by some
function $\epsilon (n)$ which tends to zero as $n$ tends
to infinity.  
It is {\it secure} against Alice if, after the point at which
she is supposed to be committed, it guarantees that 
either her probability of successfully unveiling $0$
or her probability of successfully unveiling $1$ is 
less than $\epsilon'(n)$, which again should tend to zero as
$n$ tends to infinity. 
For the protocol to be {\it perfectly reliable}, it must guarantee that,
if Alice honestly follows the protocol to commit and later
unveil the bit $b$, the
probability of her unveiling being successful is one.  
For it to be {\it reliable}, it must guarantee that this 
probability is greater than $1 - \epsilon'' (n)$, where 
$\epsilon''(n)$ again should tend to zero as $n$ tends to infinity.  

There may be a case for treating reliability and
security on an equal footing.  However, the usual definition
of bit commitment requires security and perfect reliability, but 
not necessarily perfect security, and we follow this convention here, 
since the distinction makes no essential difference to 
our discussion.  That is, we require that a bit commitment
protocol should be secure against both parties and perfectly reliable.  
If a protocol can guarantee this, 
regardless of the technology or computing power available to the parties, 
within the model defined by some physical theory, 
then it is {\it unconditionally secure} within that
theory.  The relevant physical theory here ---
until we consider the role of quantum information, in 
section \ref{quantumsecurity} onward --- is relativistic
classical physics in Minkowski space.

We use units in which the speed of light $c=1$ and choose
inertial coordinates, so that the minimum possible time
for a light signal to go from one point in space to another
is equal to their spatial separation.  
We consider a cryptographic scenario in which coordinates
are agreed by Alice and Bob, who also agree on two points 
$\ul{x}_1, \ul{x}_2$.  Alice and Bob are 
required to erect laboratories, including sending and receiving
stations, within an agreed distance $\delta$ of the points, where 
$ \Delta x = | \ul{x}_1 - \ul{x}_2 | \gg \delta $.  
These laboratories need not be restricted in
size or shape, except that they must not overlap. 

We refer to the laboratories in the vicinity of $\ul{x}_i$ as 
$A_i$ and $B_i$, for $i = 1$ or $2$. To avoid unnecessarily
proliferating notation, we use the same labels
for the agents (sentient or otherwise) assumed to 
be occupying these laboratories.  
The agents $A_1$ and $A_2$ 
may be separate individuals or devices, but we assume 
that they are collaborating with complete mutual trust and with
completely prearranged agreements on how to proceed, to the extent 
that for cryptanalytic purposes
we can identify them together simply as a single entity,
Alice ($A$); similarly $B_1$ and $B_2$ are identified as Bob ($B$).  

As usual in defining a cryptographic scenario for a protocol
between mistrustful parties, we suppose
Alice and Bob each trust absolutely the security and integrity of their own
laboratories, in the sense that they are confident that 
all their sending, receiving and analysing devices 
function properly and also that nothing within their laboratories
can be observed by outsiders.  They also have confidence in the
locations of their own laboratories in the agreed coordinate
system, and in clocks set up within their laboratories.    
However, neither of them trusts any third party or channel or device
outside their own laboratory.  We also assume that $A_1$ and $A_2$ 
either have, or can securely generate as needed during the protocol, an
indefinite string of shared secret random bits.   

Strictly speaking, if $A_1$ and $A_2$ are disconnected laboratories, this
arrangement requires $A$ to trust {\it something} outside
her laboratories --- for instance the coordinates of distant stars --- in
order to establish their relative locations, and similarly for $B$.
Purists in the matter of cryptographic paranoia might thus instead
require that $A_1$ and $A_2$ are separated regions within one large
laboratory controlled by $A$ (which must be long in one dimension, of
size order $\Delta x$, but could be small in the other two
dimensions), and similarly for Bob's agents, with $A$'s and $B$'s
laboratories still of course kept disjoint.  In this case, agents in
$A_1$ and $A_2$ can establish their relative separation by
measurements within $A$'s laboratory; similarly agents in $B_1$
and $B_2$ need only carry out measurements within $B$'s laboratory.  Though the
participants still need to believe that Minkowski space is a good
approximate description of the world outside their laboratories in
order to have confidence in the protocol, they then need trust nothing
outside their laboratories in order to trust their implementation of
it.  This arrangement also allows $A_1$ and $A_2$ to use a secure
channel within their shared laboratory to share secret random bits as
needed during the protocol, eliminating the need for them either to
share an unbounded string of secret random bits or to 
establish a secure external channel between two disconnected 
laboratories. 

To ensure in advance that their clocks are synchronised and that
their communication channels transmit at sufficiently near light
speed, the parties may check that test signals sent out from each of
Bob's laboratories receive a response within time $4 \delta$ from
Alice's neighbouring laboratory, and vice versa.  However, the parties
need not disclose the precise locations of their laboratories in order
to implement the protocol.  Nor need Alice or Bob take it on trust
that the other has set up laboratories in the stipulated region. 
(A protocol which required such trust would, of course, be fatally
flawed.)  The 
reason is that each can verify that the other is not significantly
deviating from the
protocol by checking the times at which signals from the other party
arrive.  For each party can verify from
these arrival times, together with the times of their own
transmissions, that particular specified pairs of signals, going from
Alice to Bob and from Bob to Alice, were generated independently
--- and this guarantee is all that is required for security.  

Given a laboratory configuration as above, one can set out
precise timing constraints for all communications in a protocol in
order to ensure the independence of all pairs of signals which are
required to be generated independently.  We may use the time coordinate in
the agreed frame to order the signals in the protocol.  (Without such a
convention there would be some ambiguity, since the time ordering is
frame dependent).  

In fact, the protocols we consider are unconditionally 
secure against Bob regardless of his actions: Alice needs no guarantees about
the location of his laboratories or 
the independence of the messages he sends during the protocol. 
However, Bob needs some guarantees about the independence of
Alice's messages, which can be ensured by the following arrangement.

We say that two spacetime regions $P$ and $Q$ are {\it sufficiently
spacelike separated} if it is the case that if Alice receives a
message sent from within the region $P$ and sends a reply which is
received by Bob within that region, and receives a message from within
$Q$ and sends a reply which is received by Bob within $Q$, then Bob
will be assured that the first reply was generated independently of
the second received message and the second reply was generated
independently of the first received message.  Now we choose two
sequences of spacetime regions $P_1 , P_2 , \ldots$ and $Q_1 , Q_2 ,
\ldots$ with the properties that:
\vskip 5pt

\noindent{\bf (i)}  $P_1$ is sufficiently spacelike
separated from $Q_1$, $Q_1$ from $P_2$, $P_2$ from $Q_2$, $Q_2$ from
$P_3$ and so on;\hfill\break 
{\bf (ii)} the spacetime regions $P_1 , P_2 , \ldots $ are
defined by the same region $P$ in space, namely the ball of radius
$\delta$ around the point $\ul{x}_1$, and successive disjoint time
intervals $[ s_1 , s'_1 ] \, , [ s_2 , s'_2 ] \, , \ldots$;
similarly the $Q_j$ are defined by the same spatial region $Q$, namely
the ball of radius $\delta$ around the point $\ul{x}_2$, and
successive disjoint time intervals $[ t_1 , t'_1 ] \, , [ t_2 ,
t'_2 ] \, , \ldots$; \hfill\break
{\bf (iii)} the regions are strictly time ordered in
the agreed time coordinate, with ordering $P_1 \, , Q_1 \, , P_2 \, ,
Q_2 \, , \ldots $, so that we have $s_1 < s'_1 < t_1 < t'_1 < s_2 <
s'_2 < t_2 < t'_2 < \ldots$; \hfill\break
{\bf (iv)}  the time intervals are all 
of the same length $\Delta t$, where $\Delta t > 4 \delta$
and $\Delta t \ll \Delta x$. \hfill\break
\vskip5pt
Note that these conditions allow a
single agent (person or device) $A_1$ to be responsible for all
Alice's communications from the regions $P_i$ and a second agent $A_2$
to be responsible for all her communications from the regions $Q_i$,
and similarly for Bob, as suggested by the earlier discussion.

It is then easy to define timing constraints, in terms of 
$\delta \, , \Delta x$ and $\Delta t$, which, if
respected by $A$ and $B$, and if $A$'s and $B$'s laboratories are
sited as prescribed, ensure that $A_1$ can receive a message from
$B_1$ sent from within $P_1$ and send a reply which will be received
within $P_1$, then that $A_2$ can receive a message from $B_2$ sent
from within $Q_1$ and send a reply which will be received within
$Q_1$, then that $A_1$ can receive a message from $B_1$ sent from
within $P_2$ and send a reply which will be received within $P_2$, and
so on.  
Each message-reply exchange constitutes one {\it round}
of the commitment protocol.  To keep the notation simple in
the following discussion, we take 
these constraints as implicitly specified, and identify rounds
of the protocol by the spacetime region $P_i$ or $Q_i$ in
which they take place. 
 
\section{The relativistic bit commitment protocol $RBC1$}

We now recall the bit commitment protocol --- call it $RBC1$ ---
described in Ref.~\cite{kentrel}. 

Alice and Bob first agree a large number $N$, a security parameter
for $RBC1$.  
In what follows we take $N = 2^m$ to be a power of $2$,
and where convenient also refer to $m$ as the security parameter. 
Taking $N$ to be a power of $2$ simplifies the 
description of the protocol, since it
allows the various random numbers generated and transmitted
by the parties to be efficiently coded in binary.  
Note, however, that both $RBC1$ and $RBC2$ (defined below) can be defined 
for any $N$.  Optimally efficient implementations --- those which attain
a given security level with minimal communication requirements ---
may generally require $N$ to be other than a power of $2$. 

All arithmetic in the protocol is carried out modulo $N$. 
Before the protocol begins,
$A_1$ and $A_2$ agree on a list $\{ m_1 , m_2 , \ldots \}$
of independently chosen random integers in the range $0 \leq m_i < N$.  
$B_1$ and $B_2$ also need to generate lists of random pairs of integers
$(n_{j,0} , n_{j,1})$ in the range $0 \leq n_{j,0} , n_{j,1} < N$; 
these numbers, which need not be agreed between the $B_i$ in advance, 
are drawn from independent uniform distributions, with the 
constraint that $n_{j,0} \neq n_{j,1}$ for each $j$. 

In the first round of the protocol, which takes place within
$P_1$, $B_1$ sends
$A_1$ the labelled pair $(n_{1,0} , n_{1,1} )$. 
On receiving these 
numbers, $A_1$ returns $n_{1,b} + m_1$ in order to commit the bit $b$.
This completes the commitment phase.  

The second and later rounds of the protocol constitute the 
sustaining phase.  
In the second round, which takes place within $Q_1$,
$B_2$ asks $A_2$ to commit to 
him the binary form $a^1_{m-1} \ldots a^1_0 $ of $m_1$.
This is achieved by sending $A_2$ the labelled list of $m$  pairs 
$(n_{2,0} , n_{2,1} ), \ldots , (n_{m+1,0} , n_{m+1,1} )$.
$A_2$  returns $n_{2, a^1_0} + m_2 , \ldots , 
n_{m+1, a^1_{m-1}} + m_{m+1}$.

In the third round, which takes place within $P_2$, 
$B_1$ asks $A_1$ to commit in similar fashion the binary
forms of the random numbers $m_2 , \ldots , m_{m+1}$ used 
by $A_2$; in the fourth round, which takes place within $Q_2$, 
$B_2$ asks $A_2$ to commit the
binary forms of the random numbers $m_{m+2}, \ldots , m_{m^2 + m + 1}$
used by $A_1$ in the third round commitment.  
And so on; communications are continued indefinitely in order
to sustain the commitment, unless and until Alice chooses to unveil
the committed bit or either party chooses to abandon the protocol.   

In fact, the protocol's security is ensured provided that the
differences $n_{j,1} - n_{j,0}$ are random elements of 
$\{ 0 , \ldots , N-1 \}$, even if the 
integers $n_{j,0}$ or $n_{j,1}$ are not. 
A natural alternative convention would thus
be to choose $n_{j,0} = 0$ for all $j$
and take the $n_{j,1}$ to be independently randomly chosen in the 
range $0 < n_{j,1} < N$.  
If $A$ and $B$ adopt this convention, there is obviously no need 
for $B$ to transmit the values of $n_{j,0}$.  We have kept the 
option of general $n_{j,0}$ partly for notational convenience,
but also because we have no proof that every possible level
of security is most efficiently attained by setting $n_{j,0} = 0$. 

We now define the unveiling protocol.  
Define the index $\bar{\imath}$ to be the alternate value to $i$;
e.g. $A_{\bar{1}} = A_2$.  
Either or both of the $A_i$ may choose to unveil the originally
committed bit. 
For $A_i$ to unveil, she
reveals to $B_i$ the set of random numbers used by
$A_{\bar{\imath}}$ in $A_{\bar{\imath}}$'s last set of commitments, 
sending the signal sufficiently early that when $B_i$ receives it he will
be guaranteed that it was generated independently of the 
message sent by $B_{\bar{\imath}}$ to $A_{\bar{\imath}}$ to initiate 
this last set of commitments. 
To check the unveiling, $B_i$ sends the unveiling data
to $B_{\bar{\imath}}$, who checks through the data to ensure that
all the commitments in the protocol are consistent and correspond to a 
valid commitment of a bit $b$ in the first round.  If so, he accepts
that Alice was genuinely committed to the bit $b$ from the point at
which the first round was completed. 

To allow $A_1$ to unveil soon after the first round --- which
the protocol $RBC2$, discussed 
below, requires --- we need to vary this procedure,
since there is no previous round of commitments about which
she can supply data.   $A_1$ can unveil after the first round
by revealing to $B_1$ the set of random numbers which 
will be used by $A_{2}$ in the second round, 
sending the signal sufficiently early that when $B_1$ receives it he will
be guaranteed that it was generated independently of the 
messages sent by $B_{2}$ to $A_{2}$ to initiate 
the second round of commitments.

Since it takes a finite time for the relevant signals
to reach Bob's agents, the protocol is not immediately
completed by Alice's unveiling message. 
It is the need to wait for receipt of information that is unknown to 
the unveiler $A_i$ (since it depends on the last set of pairs 
sent by $B_{\bar{\imath}}$) and to the unveilee $B_i$ (since it includes
the last set of commitments sent by $A_{\bar{\imath}}$) which ensures that the 
protocol is not vulnerable to a Mayers-Lo-Chau quantum 
attack. 

$RBC1$ requires the $A_i$ and $B_i$ to exchange sequences of 
strings of exponentially increasing length, 
each exchange taking place within a time interval of length
less than $ \Delta t$.   This requires exponentially increasing
communication rates.   
It also requires either that the $A_i$ 
previously generated a list of shared secret random numbers whose length
depends exponentially on the duration of the protocol, 
or else that they generate and securely share such a list
during the protocol.  The second option requires an exponentially
increasing secure communication rate, and also requires the
generation of random numbers at an exponentially increasing rate.  
The $B_i$ also need to
generate a sequence of random numbers whose length depends
exponentially on the duration of the protocol, though they do not 
need to share them.
In the following sections we define a refinement of $RBC1$ that 
avoids all these impractical features.   

It is perhaps worth noting here that the arrangement of $A_1$
and $A_2$ used above for $RBC1$, and in a later section for
the improved protocol $RBC2$, represents just one possible
configuration of agents.  It might sometimes be useful in
practice to have more agents participating in the protocol,
or to allow the agents to be more mobile, or both.  
For instance, one might imagine
applications in which it is important for $A$ to be able to
allow any of her agents, anywhere, to initiate an unveiling 
as soon as they learn some critical fact.  One way of doing
this is to use the above arrangement for the commitment
and sustaining phases of the protocol, but to arrange in
addition that $A$ and $B$ maintain  
large numbers of additional agents densely 
distributed in the same spatial region, with all Alice's
random bits shared among all her agents.   

\vskip10pt
\section{Rudich's scheme for linking committed bits} 

In the following section we improve $RBC1$ using a technique
developed (for a different purpose) by Rudich \cite{rudich}.
Since Rudich's idea applies to any type of classical bit commitment,
it is most naturally described abstractly --- so, in this section, 
we need not consider separated laboratories, relativistic 
signalling constraints, timings or locations.   

Suppose that $A$ and $B$ are equipped with a black box oracle that 
generates secure commitments to $B$ of bits specified by $A$, and
unveils the bits to $B$ if (and only if) $A$ requests it to. 
Suppose now that $A$ wishes to commit to two bits, 
$b_1$ and $b_2$, in a way that will allow her subsequently to
demonstrate to $B$ that the committed bits are equal, without
giving $B$ any information about their shared bit value $b = b_1 = b_2$, 
and while still retaining a commitment of this shared bit $b$, which can
subsequently be unveiled if she chooses.   
Rudich some time ago \cite{rudich} pointed out a simple and elegant way to
achieve this.  A version of Rudich's scheme, slightly modified to 
adapt it for relativistic bit commitments, follows.     

First, $A$ and $B$ agree on the value of a large integer $M$, which serves
as a security parameter.  
Now, $A$ makes a {\it redundant commitment} of the bit $b_1$ 
using $2 M$ elementary bit commitments of bits $b^{ij}_1$, where
the index $i$ runs from $1$ to $2$ and $j$ runs from $1$
to $M$, with the property that the $b^{1j}_1$ are chosen
randomly and independently and 
the $b^{2j}_1$ are defined by the constraint that 
$b^{1j}_1 \oplus b^{2j}_1 = b_1$ for each $j$.   
Alice also makes a redundant commitment of the
bit $b_2$ using $4 M$ elementary commitments of 
bits, $b^{ij}_2$, defined similarly but with $j$ running from $1$
to $2M$, with the  
$b^{1j}_2$ chosen randomly and 
independently of the $b^{1j}_1$ as well as each other.  

To test the equality of the committed bits $b_1$ and $b_2$, $B$ 
proceeds as follows. 
$B$ chooses a random one-to-one map $f$ from $\{ 1 , \ldots , M \}$
to  $\{ 1 , \ldots , 2 M \}$.  
For each $j$ from $1$ to $M$, $B$  then asks $A$ whether the pairs 
$(b^{1j}_1 , b^{2j}_1 )$ and $(b^{1 f(j)}_{2} , b^{2 f(j)}_{2} )$
are equal (i.e.  $b^{1j}_1 = b^{1 f(j)}_{2}$ and  $b^{2j}_1 = 
b^{2 f(j)}_{2} $) 
or opposite (i.e.
$b^{1j}_1 = 1 - b^{1 f(j)}_{2}$ and  $b^{2j}_1 = 1 - 
b^{2 f(j)}_{2} $).   If $A$ has followed the protocol correctly,
and the bits $b_1$ and $b_2$ are indeed equal, then one of these two cases
must apply for any given $j$, and $A$ states which, for each $j$. 
$B$ then tests $A$'s answers by choosing further independent random numbers
$m(j) \in \{ 1, 2 \}$ for each $j$ from $1$ to $M$
and asking $A$ to unveil the two bits
$b^{m(j) j }_1 $ and $b^{ m(j) f(j) }_2$.   $A$ does so, and
$B$ checks that these bits are indeed equal, or opposite, as
claimed.  

If $A$ passes all these tests, $B$ accepts that indeed
$A$ was committed to two bits $b_1$ and $b_2$ with $ b_1 = b_2$.  
$B$ accepts also that the $M$ remaining unopened pairs of
bits $ (b^{1k}_{2} , b^{2k}_{2 } )$, corresponding to values $k$ not in
the image of $f$, together constitute a redundant
commitment to the common bit value $b = b_1 = b_2 $ of the 
same form as the original commitment to $b_1$.

The following security features are sufficient for our
purposes.  

{\bf Security against $B$:} \qquad Clearly, if the protocol is
correctly followed by $A$, $B$ obtains no information about $b$.  

{\bf Security against $A$:} \qquad   
The following definitions assume a fixed parameter $\gamma$.
Any value of $\gamma$ in the range $ 0 < \gamma < 1/2 $ suffices
to prove the security of Rudich's linking scheme.  
For the moment we will not specify $\gamma$ further;
we will make a specific choice later.  
We say $A$ is {\it effectively 
R-committed} (R here stands for Rudich) to $b_1$ if at least $ (1 - \gamma ) M$ of the $M$ pairs 
that are supposed to define the redundant commitment to 
$b_1$ are of the correct form: i.e. if 
$b^{1j}_1 \oplus b^{2j}_1 = b_1$ for at least $ ( 1 - \gamma ) M$ 
different values of $j$.   
We say $A$ is {\it effectively 
R-committed} to $b_2$ if at least $ (2 - \gamma ) M $ of the $2 M$ pairs 
that are supposed to define the redundant commitment to 
$b_2$ are of the correct form. 

Note that, according to these definitions, if $A$ is effectively
R-committed to $b_1$ by $M$ pairs of bit commitments
and effectively R-committed to $b_2$ by $2M$ pairs, then, after
the tests (and regardless of their result), the 
remaining $M$ unopened pairs will again effectively 
R-commit her to $b_2$ in the first sense.  That is, at least
$(1  - \gamma )M$ of the remaining $M$ pairs define a redundant
commitment to $b_2$.    

{\bf Lemma 1} \qquad 
Suppose $A$'s Rudich linking does not effectively R-commit
her to both $b_1 = b$ and $b_2 =b$, for some bit value $b$.
(That is, either she is
not effectively R-committed to at least one of the
bits, or she is effectively R-committed to 
two different bit values.)   
Let $\epsilon (M)$ be the 
maximum probability of $A$'s passing $B$'s tests, optimised
over all bit configurations satisfying these constraints.  
Then $\epsilon (M) \approx \exp ( - C M )$ as $M$ tends to 
infinity, for some positive constant $C$.  

\proof 

Suppose $( 1 - \gamma_1 )M$ of the $M$ pairs defining $A$'s first
Rudich commitment define the bit $b_1 = b$, and the remaining $\gamma_1 M$
define the bit $b_1 = \bar{b}$. 
Suppose $( 2 - 2 \gamma_2 )M$ of the $M$ pairs defining $A$'s second
Rudich commitment define the bit $b_2 = b$, and the remaining $2 \gamma_2 M$
define the bit $b_2 = \bar{b}$. 
As $A$ is not effectively R-committed to either $b$ or $\bar{b}$ with
both commitments, we have that $\max( \gamma_1 , 2 \gamma_2 ) > \gamma $ and
$\max ( 1 - \gamma_1 , 2 - 2 \gamma_2 ) > \gamma$.

Each of the $\gamma_1 M$ bits from the first commitment 
that are committed to $\bar{b}$ has a probability $(1 - \gamma_2 )$ 
of being tested against a commitment to $b$ from the second commitment,
Each of the $( 1 - \gamma_1 ) M$ bits from the first commitment 
that are committed to $b$ has a probability $\gamma_2 $ 
of being tested against a commitment to $\bar{b}$ from the second commitment.  
The probability of passing all these tests is of order 
$$
\bigl( \frac{1}{2} \bigr)^{ M ( \gamma_1 ( 1 - \gamma_2 ) + \gamma_2 (1  -  \gamma_1 ) ) } 
\leq 
\bigl( \frac{1}{2} \bigr)^{  \frac{M \gamma}{2} } \, .
$$ 
(The inequality follows since $ x ( 1 - y ) + y (1 -x ) \geq \frac{\gamma}{2}$
for any $x,y$ in the range $(0,1)$ such that 
$\max(x,y) > \frac{\gamma}{2}$ and $
\max(1-x,1-y) > \frac{\gamma}{2}$.) 

\remark  To keep a sequence of linked bit commitments secure against
$A$, $B$ wants to ensure that he will almost certainly detect cheating
unless $A$ is effectively R-committed to the same bit for the first
and last commitments.  From the note preceding the above lemma, 
we see that, if $A$'s first and last commitments 
are not effectively R-committed to the same bit value, 
some intermediate adjacent pair of commitments must 
fail to be effectively R-committed.  
Any value of $\gamma < \frac{1}{2}$ such that $\gamma M$ is 
an integer can be used in the definition of effective 
R-commitment here.  We can now apply the lemma and minimise 
our bound on the probability
of $A$ escaping detection by taking 
$\gamma M  = \lfloor \frac{M-1}{2} \rfloor$,
where $\lfloor x \rfloor$ is the largest integer less than or equal to $x$.  
We thus see $A$'s probability of successful cheating is bounded
by a term of order
$$ ( \frac{1}{2} )^{ \frac{1}{2} ( \lfloor \frac{M-1}{2} \rfloor ) } 
\approx ( \frac{1}{2} )^{ \frac{M}{4} }  \, .
$$

\vskip 10pt
\section{Use of Rudich's linking in finite channel relativistic bit commitment} 

Our key idea for improving $RBC1$ is the following.   
If $b_1$ represents a relativistic bit commitment, and $b_2$ another
relativistic bit commitment begun at a later round than $b_1$, then 
$A$ can use Rudich's technique to 
link these commitments by showing that $b_1 = b_2 = b$. 
After doing so, she maintains a commitment to their common
bit value, drawn from a subset of the elementary commitments
used for $b_2$.  She can thus abandon the commitment to
$b_1$ at this point, while remaining committed to $b$ via
a subset of the elementary commitments for $b_2$, and
start a new commitment for a bit $b_3$.  
Letting $b_i$ take the role of $b_{i-1}$ for $i = 2, 3$, she
then repeats the procedure above iteratively. 
$A$'s commitment to the bit $b$ is thus 
always defined by elementary commitments
made in a recent round, so avoiding the exponential blowup
that makes $RBC1$ impractical.  

There are a variety of ways of implementing this basic strategy.
One of the simplest is the following protocol, which we call $RBC2$. 
This uses iterations of the relativistic bit commitment protocol $RBC1$ as
sub-protocols.  $RBC2$ has two security parameters: $N$, the security
parameter for the $RBC1$ sub-protocols, and $M$, the security parameter
used in the Rudich linking protocol.   

The first iteration of the linking mechanism is as follows. 
In region $P_1$, 
$A_1$ redundantly commits a bit $b_1 =b$ using a Rudich coding with
$M$ pairs of individual relativistic bit commitments made using $RBC1$. 
This relativistic commitment is sustained by $A_2$ in the region
$Q_1$.  
A second redundant bit commitment, $b_2 =b$,
using $2M$ pairs of individual relativistic bit commitments, 
is begun by $A_1$ in the region $P_2$.  
Also in the region $P_2$, $A_1$ and $B_1$ go through the Rudich 
linking protocol for $b_1$ and $b_2$.  In the case of $b_1$, 
$A_1$ can unveil elementary relativistic bit 
commitments, when required by the linking protocol, by revealing the random
numbers used by $A_2$ in $Q_1$ to sustain these commitments. 
In the case of $b_2$, $A_1$'s required unveilings are made by revealing the 
random numbers that will be used by $A_2$ in $Q_2$ to sustain 
the relevant commitments.   

In both cases, $B$ needs to collect together the information 
supplied by $A_1$ (in $P_2$) and $A_2$ (in $Q_1$ or $Q_2$) 
in order to verify that $A$ has passed the tests in the 
linking protocol.  Once this is done, and $B$ has established
that $b_1 = b_2 $, the remaining unopened elementary commitments 
defining $b_1$ may be discontinued.  All the $2M$ pairs of elementary
commitments defining $b_2$ are sustained by $A_2$ in $Q_2$,
since she does not know which of the elementary commitments was
unveiled by $A_1$ in $P_2$.
Of these $2M$ pairs, $M$ remain unopened and define a redundant
commitment for $b$, and both $A_1$ and $B_1$ 
know the identity of these $M$ after communications in $P_2$ have
ended.   They can thus use the redundant bit commitment defined
by these $M$ pairs for another iteration of the linking mechanism in $P_3$.

In this second iteration, in the region $P_3$, 
$A_1$ makes a third redundant bit 
commitment, $b_3 =b$, to $B_1$, using $2M$ pairs of individual relativistic
bit commitments, and goes through a second Rudich linking, proving
(eventually) to $B_1$ that $b_3 = b_1$.    

And so on: $A_1$ initiates a new redundant commitment to, and goes
through a Rudich linking with, $B_1$, for each round they 
participate in after the first.  $A_2$ and $B_2$ simply sustain
each of the commitments initiated by $A_1$ and $B_1$ for one round.  

\section{Proof of security of $RBC2$ against classical attacks} 

{\bf Security against $B$:} \qquad  If the $A_i$ honestly 
follow the protocol using 
secret shared independently generated random 
numbers, then whatever strategy the $B_i$ use, the information 
they receive is uncorrelated with the committed bit.  
The protocol is thus perfectly secure against Bob.  

{\bf Security against $A$:} \qquad   A key constraint on
Alice is that $B_1$ may request $A_1$ to unveil any of the
individual $RBC1$ relativistic bit commitments made during the
protocol.     $B_1$ will request $M$ of the first batch of $2M$ 
commitments to be unveiled during the linking protocol in $P_2$.
Of every successive batch of $4M$ commitments, $B_1$ will request 
that $A_1$ unveil $M$ of them as soon as they are committed,
and another $M$ two rounds later.   Thus, if $M$ is large,
unless $A_1$ is able to
provide a valid unveiling of a bit for almost every commitment in every
batch, she will almost certainly be discovered to be cheating.  

Now, the relativistically enforced security of $RBC1$ means that 
$A_1$ has no strategy that will allow her a high probability
of successfully producing valid unveilings for both $0$ and
$1$ for any of the elementary $RBC1$ commitments.   
To unveil a commitment that she has just initiated in $P_i$, 
she must supply $m$ $m$-bit numbers that correspond to a 
valid commitment by $A_2$ in $Q_i$ of an $m$-bit number 
that corresponds to a valid commitment of a bit $b$ by $A_1$ 
in $P_i$.   Similarly, to unveil in
$P_{i+1}$ a commitment she initiated in $P_i$, she needs to
supply $m$ $m$-bit numbers that correspond to a valid 
commitment by $A_2$ in $Q_i$ of an $m$-bit number that
corresponds to a valid commitment of a bit $b$ by $A_1$ in $P_i$.  
Either way, if she is able, with significant probability, to produce valid 
unveilings of both values of $b$, she is also able to infer, with
significant probability, the differences between the $m$ pairs 
of random numbers, $(n_{i,0} - n_{i,1})$ sent by $B_2$ to $A_2$
in the commitment round in $Q_i$.   (With our convention that
$n_{i,0} = 0$, this is equivalent to inferring the $n_{i,1}$.)
But, if $B$ follows the protocol honestly, all $(N-1)^m$ 
possible sets of difference values are equiprobable, and 
$A_1$ can have no information available
to her, either in $P_i$ or $P_{i+1}$, about the actual
difference values, 
assuming that our current understanding of physics correctly assures us that 
superluminal signalling is impossible.  

Suppose that $N \geq 4$, so that $m \geq 2$ and $(N-1)^m \geq 9$.
Then in particular, for each elementary commitment, there can be at most 
one set of $m$ numbers that $A_1$ can use for unveiling
with the knowledge that the probability of unveiling a 
valid bit commitment is $>1/3$.  Thus, for each elementary commitment,
$A_1$ can successfully unveil at most one of the bit values $0$ and $1$ 
with probability $> 1/3$.  We say $A_1$ is {\it probabilistically
committed} to the bit value $b$ if she can successfully unveil $b$
with probability $> 1/3$.  

By analogy with our earlier definition, we say Alice is {\it probabilistically
effectively RR-committed} (RR here stands for relativistic Rudich) 
to the bit $b$ by the redundant commitment initiated
in round $1$ if she is probabilistically committed by
each elementary bit commitment in 
a subset of $(1 - \gamma)M$ of the $M$ pairs of elementary
commitments and these bit values constitute a Rudich coding for the 
bit $b$ among these pairs.  
We say she is probabilistically effectively
RR-committed to the bit $b$ by the redundant commitment initiated
in any later round if she is probabilistically committed by 
each elementary bit commitment in  
a subset of $(2 - \gamma)M$ of the $2M$ pairs of elementary
commitments and these bit values constitute a Rudich coding for the 
bit $b$ among these pairs.  

We take $\gamma < \frac{1}{2}$, so that Alice can 
be probabilistically effectively RR-committed  
to at most one bit value in any given round.  

{\bf Lemma 2} \qquad 
Suppose $A$ is not probabilistically effectively RR-committed
to both $b_1 = b$ and $b_2 = b$, for some bit value $b$,
on any two successive rounds of RBC2.  
(That is, either she is
not probabilistically effectively RR-committed to at least one of the
bits, or she is probabilistically effectively RR-committed to 
two different bit values.)   
Let $\epsilon (M)$ be the 
maximum probability of $A$'s passing $B$'s tests, optimised
over all bit configurations satisfying these constraints.  
Then $\epsilon (M) \approx \exp ( - C M )$ as $M$ tends to 
infinity, for some positive constant $C$.  

\proof 

Suppose $y_1 M$ of the $M$ pairs defining $A$'s first
Rudich commitment define the bit $b_1 = b$, and that $\bar{y}_1 M$
define the bit $b_1 = \bar{b}$. 
Suppose $y_2 M$ of the $M$ pairs defining $A$'s second
Rudich commitment define the bit $b_2 = b$, and that $\bar{y}_2 M$
define the bit $b_2 = \bar{b}$. 
Each of the $y_1 M$ bits from the first commitment that
are committed to $b$ has probability $( 1 - y_2 )$ of 
being tested against a pair from the second commitment that
does not commit to $b$.  
Similar calculations for the other possibilities lead to
an expected total of $( 1 - y_1 y_2 - \bar{y}_1 \bar{y}_2 )M$ 
tests that have probability $\geq \frac{1}{3}$ of failure.  

As $A$ is not probabilistically effectively RR-committed to either $b$ or $\bar{b}$ with
both commitments, we have that $\max( 1 - y_1 , 2- 2y_2 ) > \gamma $ and
$\max ( 1 - \bar{y}_1 , 2 - 2 \bar{y}_2 ) > \gamma$.
By an argument similar to that used in the proof of Lemma 1,
the probability of passing all the tests 
is bounded by a term of order 
$$
\bigl( \frac{2}{3} \bigr)^{ M ( \frac{1}{2} \gamma ) } \, .
$$ 
QED.

\remark Note that if $A_1$ is probabilistically 
effectively committed to a Rudich
coding for a bit $b$ by a commitment initiated in $P_i$, she cannot
be probabilistically effectively committed to a coding for the
opposite bit $\bar{b}$ in $P_{i+1}$, since she 
learns no information in the mean time
about the round that takes place in $Q_i$.  
Thus, unless she is probabilistically effectively committed to
the same bit in both $P_i$ and $P_{i+1}$, she must
fail to be probabilistically effectively committed to any
bit value in at least one of the two regions.

As before, we take $\gamma M  = \lfloor \frac{M-1}{2} \rfloor$.  
We thus see any cheating by $A$ will be detected with
probability of order
$$ \bigl( 1 - (\frac{2}{3})^{ \frac{1}{2} ( \lfloor \frac{M-1}{2} \rfloor ) } \bigr)
\approx \bigl( 1 - ( \frac{2}{3})^{ \frac{M}{4} } \bigr) \, .
$$

\remark  Alice can only usefully cheat either by failing to be 
probabilistically effectively Rudich committed to any bit by some set of
$M$ pairs at some point, or by being effectively Rudich committed
to some bit $b$ by a set of $M$ pairs and failing to be effectively
Rudich committed to $b$ by the corresponding set of $2M$ pairs at
some point.   It is not hard to show that, among the commitment
strategies covered by these options, the one that minimises her 
probability of being detected cheating in the relevant Rudich
test is for all $M$ pairs from the first set to probabilistically 
commit to the same bit $b$, while $ (\lfloor \frac{2M-1}{4} \rfloor +
1 )$ (i.e. $\frac{M}{2}$ if $M$ is even, and $\frac{M+1}{2}$ if $M$ is odd) of
the pairs from the second set commit to $\bar{b}$ and the 
remainder to $b$.   We can thus find an exact lower bound
on the probability $p(M)$ of being detected: for example, if $M$ is even
we have 
$$
\label{detectbound} 
p(M) \leq   \sum_{r=0}^{M/2} ( \frac{2}{3} )^r \left( \begin{array}{c}
  \frac{M}{2} \\ r \end{array} \right) 
\left( \begin{array}{c}
  \frac{3M}{2} \\ M - r \end{array} \right) 
\left( \begin{array}{c}
  2M \\ M \end{array} \right)^{-1}
$$

For suitably large $M$, deviating from the protocol risks detection 
while giving her 
essentially no possibility of successfully cheating. 
If Alice is rational and values her reputation for integrity, she 
will thus honestly follow the
protocol throughout, committing (not just probabilistically effectively
committing) herself to a Rudich coding of the same bit value $b$ in each round.   

\section{Definitions of security for quantum bit 
commitment}\label{quantumsecurity}

We want to consider the security of $RBC2$ when the 
possibility of the parties sharing, storing and manipulating
quantum information is taken into account.  We still assume
that physics takes place in Minkowski spacetime, with the
Minkowski causal structure, but now suppose that the correct
description of physics is some relativistic version of 
quantum theory.  Without specifying the details of this
theory, we simply assume that the 
parties can devise their own consistent labelling of
physically realisable orthogonal basis states.
To allow as much scope for cheaters as possible, we also
assume that any localised party is able to apply arbitrary local 
quantum operations to quantum states in their possession.  

Before considering the specifics of $RBC2$, we need to consider
the general definition of security for a bit commitment protocol
that may involve quantum information.  
The definitions of reliability and perfect reliability
and of security and perfect security against Bob 
are as for classical bit commitment.
However, we need a more general definition of security against Alice.
Suppose that we have reached
the first point in the protocol at which Alice is supposed to be 
committed.   We define $p_0^{\rm sup}$ to be
supremum of the set of her
probabilities of successfully unveiling $0$ over
all strategies she could pursue from this point onwards;
similarly $p_1^{\rm sup}$.  
We say that the protocol is {\it perfectly secure}
against Alice if, for any possible
initial commitment strategy, it guarantees that
$p_0^{\rm sup} + p_1^{\rm sup} \leq 1$.  
It is {\it secure} against Alice if, for any possible
initial commitment strategy, it guarantees that 
$p_0^{\rm sup} + p_1^{\rm sup} \leq 1 + \epsilon'(n)$, 
where $\epsilon' (n)$ tends to $0$ as $n$ tends to 
infinity \cite{kentbccc}.

To see the point of the definitions, consider the 
standard model of classical bit commitment in which
Alice writes her committed bit on a piece of paper
and places it in a safe, which she then locks, before
handing the safe over to Bob.  In this model, she
unveils the bit by giving Bob the combination, 
allowing him to open the safe and read the paper.
Modulo physical assumptions about the impossibility
of remotely manipulating the contents of the safe, 
this is perfectly secure against Alice.  

Now, in an idealised quantum version of this protocol, Alice could
create a superposition $ \alpha \ket{0}_A \ket{0}_S + \beta \ket{1}_A
\ket{1}_S$, where $\ket{0}_S$ and $\ket{1}_S$ are the commitment
states for $0$ and $1$ respectively, and then place the second system,
described by the $\ket{}_S$ states, in the safe.  The probabilities of
Bob reading the commitments for $0$ and $1$ when he opens the safe are
respectively $ | \alpha |^2 $ and $ | \beta |^2 $, which sum to $1$.

Modulo the same physical assumptions as before, this is a perfectly
secure protocol that commits Alice, in the sense that she cannot
alter the contents of the safe or affect the probabilities of $0$ and
$1$ being unveiled.  However, it does not satisfy the definition
of perfect security we gave for classical bit commitment, which
requires that either the probability of successfully unveiling $0$ or
that of successfully unveiling $1$ should vanish.  
This distinction between classical and quantum definitions 
of security was noted and discussed in Ref.~\cite{bcms}. 

This suggests that the natural definition for perfect security
for a quantum bit commitment protocol 
is that given above, while the classical definition
implies something stronger, which we call
{\it bit commitment with a certificate of
classicality} \cite{kentbccc}.   The quantum definition of 
imperfect security is then a natural extension: as in the
classical case, it allows that Alice may have scope for cheating 
but requires that any cheating advantage can be made
arbitrarily small.   

Another historical reason for favouring 
the definitions of quantum security just
given, and against applying the classical definitions to quantum
bit commitment, is that the former, unlike the latter, are in line
with Mayers' and Lo-Chau's analyses of cheating strategies for
quantum bit commitment protocols.  Mayers' and Lo-Chau's essential
results were that, in non-relativistic quantum bit commitment protocols
conforming to their cryptographic models, perfect security against
Bob implies that $p_0^{\rm sup} + p_1^{\rm sup} = 2$, and that in 
a protocol with security parameter $n$, security
against Bob implies that $p_0^{\rm sup} + p_1^{\rm sup} = 2 - \delta (n)$, 
where $\delta (n)$ tends to $0$ as $n$ tends to infinity. 
In other words, all such quantum bit commitment protocols are 
insecure (in fact, maximally or near-maximally insecure) 
by our quantum definition.  On the other hand, if we
regarded a certificate of classicality as part of the
definition of a quantum bit commitment, showing that unconditionally
secure quantum bit commitment is impossible would only require showing that 
no quantum protocol can guarantee to prevent the commitment of superposed
bits --- a much simpler result that can be proven in one line,
without appealing to Mayers' and Lo-Chau's analyses \cite{kentshort}.      

\section{Security of $RBC2$ against quantum attacks}

{\bf Security against $B$:} \qquad  Allowing for the possibility
of Bob storing or manipulating quantum information does not
affect the security argument given earlier.  If the $A_i$ honestly 
follow the protocol using secret shared independently generated random 
numbers, then whatever strategy the $B_i$ use, the information 
they receive is uncorrelated with the committed bit.  
The protocol is thus perfectly secure against Bob.

{\bf Security against $A$:} \qquad Alice certainly has the
option, allowed by our definition of security for quantum
bit commitment, of committing a qubit belonging to
an entangled superposition.  To see this, consider first a quantum strategy
applicable to $RBC1$.  Suppose that before the protocol
$A_1$ and $A_2$ share a commitment state $ \alpha \ket{0}_1 \ket{0}_2 + 
\beta \ket{1}_1 \ket{1}_2$ and a string of 
superposed number states of the form $\sum_{r=0}^{N-1} \ket{r}_1 \ket{r}_2$.
If $A_1$ receives the pair $(n_{1,0} , n_{1,1} )$ from $B_1$
and runs the first round of the protocol on a quantum computer, 
using her part of the commitment state and the first
superposed number state as input, and returns $k$ as output, 
the joint state becomes
$$
\alpha \ket{0}_1 \ket{k - n_{1,0}}_1 \ket{0}_2 \ket{k - n_{1,0}}_2 + 
\beta  \ket{1}_1 \ket{k - n_{1,1}}_1 \ket{1}_2 \ket{k - n_{1,1}}_2 
$$
tensored with the remaining superposed number states \cite{bcms}.
$A_2$ and $A_1$ can proceed similarly at each successive round,
maintaining a superposed commitment.  Either of them can consistently 
unveil the bit when required, by measuring their
commitment state and their random number states, producing
a consistent unveiling of $0$ (with probability $| \alpha |^2$) or 
of $1$ (with probability $| \beta |^2$).  
But now it is clear that the same basic strategy extends to $RBC2$, 
provided the $A_i$ share appropriate superposed entangled states:
in particular, every unveiling required in $RBC2$ can be consistently
made without destroying the quantum superposition of the committed bit.  

Of course, this strategy of using shared entanglement 
gives Alice no control over which bit 
will be unveiled, so its existence is consistent with the 
quantum security of $RBC1$ and $RBC2$.   
To establish that security, we need arguments parallel to those 
in the classical case, establishing that the $A_i$ 
have essentially no alternative but to implement this 
strategy honestly, since any cheating will be detected with
near certainty.  We conjecture that this is indeed the case,
and that both protocols are secure against quantum attacks.

One can easily show that $RBC1$ and $RBC2$ are temporarily
secure against quantum attack, round by round, in the following sense.  

{\bf Lemma 3} \qquad 
Let $p_0^{j, \rm sup}$ be the 
supremum of the set of probabilities of 
the relevant $A_i$ successfully unveiling $0$ in
round $j$, for all possible unveiling strategies she could 
implement in that round; similarly $p_1^{j, \rm sup}$.  
Then $p_0^{j, \rm sup} + p_1^{j, \rm sup} \leq 1 + \epsilon$, 
where $\epsilon \equiv \epsilon (N)$ (in the case of $RBC1$)
or $\epsilon \equiv \epsilon (M,N)$ (in the case of $RBC2$),
and in either case tends to $0$ as the security parameter(s)
tend(s) to infinity.  

{\bf Proof} \qquad  First note that at any given point in the protocol
$A_1$ and $A_2$ share a quantum state.  We may suppose without
loss of generality that, 
whenever the $B_i$ receive a state from the $A_i$ during the
protocol, they measure it in the computational basis.  
The $A_i$ thus share no entanglement with the $B_i$.
Hence, if we include in the definition of their shared
state all ancillae that the $A_i$ 
may have ready for use later in the protocol (either for
continuing the commitment or for unveiling), then without loss of 
generality we can take their shared state to be pure: call it $\ket{\psi}$. 

Let $A_i$ be the party attempting an unveiling on round $j$. 
To simplify the discussion a little we
assume that the suprema defined above are attainable: that is,
optimal strategies exist.  (The argument below can easily be
extended to cover the possibility that the suprema are not
attained, by considering near-optimal strategies.)  

Her optimal strategy
for attempting to unveil a $0$  --- the optimal strategy that she can 
construct given the knowledge available to her, that is --- must be 
defined by a projective decomposition of the identity $\{ A_i \}_{i=0}^m$,
representing a von Neumann measurement she will carry out on
$\ket{\psi}$, together with an assignment of distinct lists of numbers
$\underline{r_i}$ to each of the $A_i$.  (We can represent general
measurements in this way, as we allowed ancillae 
to be included in the definition of $\ket{\psi}$.) 
Her unveiling will then consist
of carrying out the measurement, and announcing the $\underline{r_i}$ 
corresponding to the result $A_i$.  At most one of the $\underline{r_i}$ 
can correspond to a valid unveiling of $0$.  Without loss of generality
we can assume that precisely one of them does --- otherwise there
is nothing to prove --- and let it be $r_0$.   

Similarly, her optimal strategy for attempting to unveil a $1$ must
be defined by a projective decomposition of the identity $\{ B_i \}_{i=0}^n$,
together with an assignment of distinct lists of numbers
$\underline{s_i}$ to each of the $B_i$. At most one of 
the $\underline{s_i}$ can 
correspond to a valid unveiling of $1$.  Without loss of generality
we can assume that precisely one of them does --- otherwise there
is nothing to prove --- and let it be $\underline{s_0}$.   

Note that we make no assumption here about the relation between the
$A_i$ and $B_i$: in particular, we do not assume that they represent
the same decomposition or commuting decompositions.

We have that
$$p_0^{j, \rm sup} = | A_0 \ket{\psi} |^2 \, , \qquad
p_1^{j, \rm sup} = | B_0 \ket{\psi} |^2 \, . 
$$
But now 
$$ | A_0 B_0 \ket{ \psi } |^2 \geq ( | A_0 \ket{\psi} | - 
   | ( 1 - B_0) \ket{\psi} | )^2  = 
( ( p_0^{j, \rm sup} )^{1/2} - ( 1 - p_1^{j, \rm sup} )^{1/2} )^2 \, . 
$$
Hence if $(p_0^{j, \rm sup} + p_1^{j, \rm sup} - 1)$ 
is significantly positive, there is a strategy available to $A_i$ 
--- applying the $B$ projections followed by the $A$ projections ---
that has a significant probability of yielding a valid unveiling for
both $0$ and $1$.   This means that, with significant probability,
$A_i$ learns information that reveals the differences 
$n_{k,1} - n_{k,0}$ in the random numbers sent to $A_{\bar{\imath}}$
in the previous round.  But it is impossible for her to obtain
any information about these differences at this point, since
a light signal cannot yet have reached her.  QED

Lemma $3$ shows in particular that RBC1 
and RBC2 are not vulnerable to the 
type of attack shown by Mayers and Lo-Chau to
imply the insecurity of non-relativistic quantum bit
commitment schemes.  In a Mayers-Lo-Chau attack on a 
protocol perfectly secure against Bob, Alice can successfully
unveil either $0$ or $1$, each with probability one.   
This is impossible in RBC1 and RBC2, essentially because
the Mayers-Lo-Chau attack would require Alice to implement
one of two unitary operations when unveiling, and, while
both operations do indeed exist, she would need to know 
the random data supplied by Bob on the previous round
in order to be able to construct both of them simultaneously.  

A full quantum security analysis would need to consider
all possible quantum operations that Alice might perform.
In general, the $A_i$ may initially share arbitrarily 
many entangled states of their choice, they may generate and 
share further entangled states during the
protocol, and they may also communicate classically during the
protocol.  Whenever one of them is required to 
respond to Bob's queries, she may carry out arbitrary quantum operations 
and measurements on the states in her possession before doing so.
These quantum operations may depend on all communications previously 
received from Bob or from her partner.  We conjecture that
RBC1 and RBC2 are secure against general quantum attacks, 
but have no proof at present.  

\section{Comments on practicality} 

Security definitions need only consider the asymptotic behaviour of
a protocol, but real world implementations require 
finite values of the security
parameters.   Here we give some estimates of the 
degree of security attainable against classical attacks,
for realistic security parameter choices. 
From these we can deduce what is needed in practice to 
implement $RBC2$ with near-perfect security against 
classical attacks.  

$RBC2$ requires $A_2$ to maintain $4M$ individual relativistic
bit commitments on each round.  Sustaining each of these
commitments requires her to commit $m$ bits, and each of
these last commitments requires $m$ bits to be transmitted.
Her total bit transmission rate is thus $4 M m^2$ per round.
To allow these commitments, $B_2$ needs to send her $4 M m^2$ bits per 
round (we assume here that they use the convention $n_{i,0} = 0$,
saving $B_2$ a factor of $2$).  These communications
thus take a total of $8 M m^2$ bits per round.

On the third and later rounds, $A_1$ needs to send $4Mm$ bits to
initiate $4M$ commitments,  
$M$ bits to respond
to $B_1$'s queries about pair relations during the linking tests,
and $2 M m^2 $ bits for the unveiling required in the Rudich
linking subprotocol.  
$B_1$ needs to send $4M m$ bits to initiate the commitments
(again assuming the convention $n_{i,0}=0$), 
send $\approx M \log_2 M$ bits to make a random choice 
among the $ \frac{(2M)!}{M!}$ possible linkings of
pairs, and a further $2M$ bits to query the pair relations.
These communications thus take a total of 
$\approx M ( 2 m^2 + 8 m + \log_2 M + 3)$ 
bits per round.  

For $m=2$, the maximum communication cost per round is
thus approximately 
$\max ( 32 M, M (27 + \log_2 M ) )$.
For small $m$ and $M$ we can easily calculate
the communication cost precisely, as we do in the following
illustrative calculations.  

We assume a $10$ GHz transmission rate 
and require each round to take place within a tenth of the time 
it would take light to travel between $A_1$ and $A_2$. 
We consider two possibilities\vskip 5pt
\noindent{\bf Case I:} $m=2$ and $M=40$;\hfill\break
\noindent{\bf Case II:} $m=2$ and $M=200$.\hfill\break
\vskip 5pt

For case I, the maximum communication cost per round
is $1280$ bits, requiring $ 1.3 \times 10^{-7}~{\rm
  sec}$.  Conservatively allowing a 
factor of $3$ to include the $A_i$'s data processing time, we obtain a 
total elapsed time of $4 \times 10^{-7}~{\rm sec}$ per round.  
This allows a separation of $4 \times 10^{-6}$ light seconds,
about $1.3~{\rm  km}$.  

For Case II, the maximum communication cost is $7041$ bits 
per round.  Calculating similarly, we obtain  
separation of about $9~{\rm km}$.  

To get an indication of the security levels attained, we use the
bound derived in section \ref{detectbound} 
on the probability $p(M)$ of Alice's
cheating being detected.
This gives bounds on Alice's successful cheating probability 
of $2.4 \times 10^{-2}$ in Case I and $ 7.3 \times 10^{-9}$ in Case II.  

\section{Discussion}

The secure relativistic protocol $RBC2$ requires the two 
parties each to maintain two dedicated 
separated secure sites.  We estimate that a separation of 
around $10$km should be adequate even if near-perfect 
security against cheating is required.  
Our security analyses suggest that 
$RBC2$ can be implemented with cheating probability
bounded by roughly $2 \times 10^{-2}$ 
using roughly $ 10^3$ bits of communication per
round and by roughly $7 \times 10^{-9}$ using roughly $10^4$ bits
of communication per round.  The probability of successful cheating cannot be
smaller than $2^{-n}$ for a relativistic protocol with 
$n$ bits of communication per round, so that these cheating 
probabilities certainly cannot be improved upon by protocols
using fewer than $6$ and $27$ bits per round respectively.  
The scope for reducing the number of bits per round (and hence
the minimum site separation) for these
security levels is thus not huge: one cannot possibly hope to
obtain better than a factor of roughly $300$, and we doubt 
that this bound is attainable.  
We cannot, though, exclude the possibility that 
an improved security analysis or an improved
protocol could allow the site separation to be reduced by a factor
of $10^1$ to $10^2$ or so.  
Any more substantial reductions would require faster communications 
technology.  The required separation is inversely proportional to the 
achievable communication rate; our estimates assumed $10$ GHz communications. 

One important caveat is that there is as yet no 
security proof against general attacks by parties 
equipped with arbitrarily powerful quantum computers.   
This is not to say that the protocol can be broken
by quantum computers.  In particular, we have shown
that the protocol is immune to the Mayers-Lo-Chau quantum
attack, which breaks all earlier attempts at 
unconditionally secure bit commitment.  Nonetheless,
a complete quantum security analysis would obviously
be highly desirable.  

A subtle point, familiar by now to most experts, but 
potentially confusing to others, is also worth reiterating here.  
Using classical bit commitment protocols as a model, one 
can define a {\it bit commitment with a certificate of classicality}
to be a perfectly secure bit 
commitment protocol that guarantees from the outset
that there is a definite classical bit value, $0$ or $1$.
Classically, bit commitment and bit commitment with a 
certificate of classicality
are identical, of course, but in the quantum domain the latter 
is a stronger primitive.  
It has been known for some time that, given a protocol
for unconditionally secure bit commitment with 
a certificate of classicality, parties who can send
and receive quantum information could implement
unconditionally secure oblivious transfer \cite{yaoot} (and 
hence unconditionally secure multi-party computation,
among other tasks).   
It is also known that neither bit
commitment with a certificate of classicality \cite{kentbccc} nor
oblivious transfer \cite{loot} can be implemented with unconditional
security, even using both quantum information and 
relativistic signalling constraints.  
Our protocols do not contradict these established results: 
as noted above, RBC1 and RBC2 implement bit commitment, but 
not bit commitment with a certificate of classicality.   

An interesting feature of our protocols is that they
achieve something that at first sight might appear to
be impossible: secure {\it deniable} bit commitment.  
Suppose that, after a certain point, Alice chooses neither to sustain
the commitment nor to unveil the bit, and that her two representatives
then exchange all the data they received during the protocol.
She can then produce two different accounts of 
her actions, consistent respectively with her having 
made and sustained commitments to zero and one respectively
during the period in which she participated in the protocol.
Of course, since she has stopped sustaining the commitment
at this point, there is no compelling reason for anyone 
to believe whichever account she produces.   
However, the accounts cannot be disproved by other parties.  
Generally speaking, deniability is useful in potentially
adverse environments, in which a party may be compelled 
to give an account of their actions and may face sanctions
if that account is inconsistent with their recorded 
behaviour during the protocol.  The possibility of deniable
bit commitment seems particularly interesting given the 
potential uses of bit commitment as a sub-protocol in 
protocols for tasks such as secure elections. 

The protocol has another interesting (and related) feature.  As we
noted above, if Alice is equipped with quantum computers,
she can use RBC2 to commit an arbitrary qubit.  If her
agents choose not to sustain the commitment after some
point, and if they are able to exchange quantum information,
they can reconstruct the committed qubit (even when it
was unknown to them --- i.e. when they do not have its 
classical description, nor any other classical or quantum
information about it).  
In terminology proposed by J\"orn M\"uller-Quade and Dominique Unruh,   
the qubit commitment is {\it retractable}:
Alice can, so to speak, take the committed qubit back from Bob
if she chooses to.  As M\"uller-Quade and Unruh have 
pointed out \cite{mupriv}, one can usefully develop an abstract
black box model of quantum bit commitments which incorporates the property of
retractability: in this model, a retractable quantum bit commitment
corresponds to Alice giving Bob an ideal safe, containing a stored
qubit, which has the property that Alice can at any time
{\it either} give Bob the
power to open the safe {\it or} repossess the safe herself and open
it.   This key insight pinpoints more precisely why 
secure oblivious transfer cannot be built from RBC2 \cite{mupriv}.
Yao's construction \cite{yaoot} requires that, if
Bob chooses a suitable random subset
of Alice's commitments and finds that the unveiled 
commitments correctly describe choices and outcomes of quantum 
measurements, he may infer that the quantum measurements
corresponding to the unopened commitments were also 
implemented irreversibly.  The retractability of 
unopened quantum bit commitments invalidates this inference.  
 
In summary, we have described a bit commitment protocol
that is unconditionally 
secure against parties equipped with arbitrarily powerful
classical computers, something not previously believed to be possible.  
The protocol has the surprising and interesting feature
that it implements deniable bit commitments.  
The result demonstrates that exploiting elementary relativistic 
signalling constraints can be a surprisingly powerful tool in
classical and quantum cryptography.  
The protocol can be easily implemented with current technology. 
It would be interesting to develop practical implementations
and to examine further the potential for useful applications.   

\vskip15pt
\leftline{\bf Acknowledgments}
\vskip10pt
The protocols described in the first version of this paper
relied on inefficient linking schemes.   
I am indebted to Claude Cr\'epeau for kindly drawing my
attention to Rudich's unpublished scheme and correctly suggesting
that it might allow a more efficient way of linking 
relativistic bit commitments.  I am also very grateful 
to Gilles Brassard for many valuable comments and
criticisms and for editorial help far beyond the call of duty.   
I would also like to thank Dominic Mayers, J\"orn M\"uller-Quade, 
Steven Rudich and Dominique Unruh for very helpful discussions.  

This work was supported in part by a Royal Society University Research
Fellowship, by the project PROSECCO (IST-2001-39227) of the IST-FET 
programme of the EC, 
and by hospitality from the Oxford Centre for Quantum Computation, the
Mathematical Sciences Research Institute in Berkeley
and the Perimeter Institute.

\section{Appendix: Comment on terminology}

The cryptographic literature on classical bit commitment includes various
different illustrations of bit commitments based on computational
or physical assumptions, which could inspire subtly different 
definitions of bit commitment.
This seems not to have 
worried classical cryptographers in the past.  After all, the 
protocols all achieve the same essential result: 
after some point, Alice is committed, and she can later unveil
if she chooses.  Singling out one particular
protocol as the unique earthly representative of the Platonic ideal
of bit commitment seems hard to justify.

However, the relation between quantum and classical cryptography
involves new subtleties, and the use of relativistic signalling
constraints in cryptography may also do so, so previously
irrelevant distinctions perhaps need to be 
considered afresh.  
We have already noted the distinction between
secure quantum bit commitment and 
secure quantum bit commitment with a certificate of 
classicality.  Another possible distinction that could be 
made \cite{mayersprivate} 
deserves discussion here.  

Some standard 
classical bit commitment schemes have the property that,  
once Alice is committed, neither party need do anything
further unless and until Alice chooses to unveil the bit.  
Consider, for example, a classical bit commitment based on a 
suitable one-way function, in which Alice commits to Bob the value
$f(x)$ of the function evaluated at some $x$, which she chose
randomly subject to the constraint that $x$ has the same parity
as the committed bit.  (Here ``suitable'' means 
that it is computationally hard to extract any information 
about the parity of $x$ from $f(x)$.)  
In this protocol, commitment is complete
once Bob has received $f(x)$, and Alice can unveil at any later
time by sending $x$.  

On the other hand, in $RBC1$ and $RBC2$
the $A_i$ and $B_i$ need to continue exchanging transmissions
indefinitely in order to sustain a secure commitment; they
stop only if and when Alice chooses to unveil the bit.
If one particularly wished to make a point of stressing this feature,
one could propose new definitions distinguishing {\it definite 
bit commitment} and {\it sustained
bit commitment}, with one-way function commitments as examples of
the former and relativistic commitments as examples of the 
latter \cite{mayersprivate}.  

It seems to me, though, that there are two persuasive arguments
against this nomenclature.
The first is historical.  Some
well-known non-relativistic bit commitment schemes also
require one or both of the parties to actively maintain the commitment
indefinitely.  For instance, in the simplest standard illustration of
bit commitment based on physical assumptions, Alice writes the bit on
a piece of paper and puts it in a sealed envelope on a table between
her and Bob, in sight of them both.  
Alice is now committed, but in order to maintain the security 
of the commitment each party now needs to watch the envelope
to ensure that the other does not interfere with it, and
they must continue doing so indefinitely unless and until
Alice chooses to unveil.  The original version of the BGKW
protocol \cite{bgkw}, in which $A_2$ is imprisoned in a Faraday cage
monitored by Bob throughout the lifetime of the commitment, has the
same feature.  To change terminology now would retroactively delegitimise 
these bit commitment protocols, relabelling them as protocols for the 
newly defined cryptographic primitive of sustained bit commitment.  

The second is a question of principle: one should not conflate
the cryptographic task to be implemented and the means of
implementation.  A cryptographic task can be defined by the inputs and
outputs into a black box version of the protocol.
In the case of
classical bit commitment, Alice inputs a bit to the box, and then 
later, if and when Alice inputs an instruction to unveil, the bit is
output to Bob.  The details of how security is ensured or maintained
do not enter into this definition. 

There are many different ways of
implementing most cryptographic tasks, and setting out the
details of any given implementation has to date been considered as
properly forming part of the definition of a protocol, rather than part of
a definition of the task.  To abandon 
this fundamental distinction now would requiring rewriting
cryptological textbooks' discussions of many other primitives besides
bit commitment.  This would remove the clear and useful distinction between 
task and protocol in current nomenclature, without which 
every different protocol might be described as an instance of a 
different task.  It seems to me far 
better not to trespass onto this slippery slope.   

%
%

\end{document}